\documentclass[onecolumn,showpacs,preprintnumbers,amsmath,amssymb,floatfix]{revtex4}

\usepackage{graphicx}% Include figure files
\usepackage{dcolumn}% Align table columns on decimal point
\usepackage{bm}% bold math

\newcommand{\pom}{\tt I\! P}
\newcommand{\beq}{\begin{equation}}
\newcommand{\eeq}{\end{equation}}

\def \pom {{I\!\!P}}

\begin{document}

\title{Diffractive photon production at the LHC}
\author{C. Brenner Mariotto$^{a}$ and V. P. Goncalves$^{b}$}

\affiliation{
$^a$ Instituto de Matem\'atica, Estat\'{\i}stica e F\'{\i}sica, Universidade Federal do Rio Grande\\
Av. It\'alia, km 8, Campus Carreiros, CEP 96203-900, Rio Grande, RS, Brazil\\
$^b$ Instituto de F\'{\i}sica e Matem\'atica, Universidade Federal de Pelotas\\
Caixa Postal 354, CEP 96010-900, Pelotas, RS, Brazil
}

\begin{abstract}
In this paper we study  the photon production in single and double diffractive processes considering the Resolved Pomeron model. We estimate the rapidity and transverse momentum dependence of the cross section for the diffractive double photon and photon+jet production. A comparison with the inclusive production is presented. We predict large values for the total cross sections, which makes the experimental analysis of these observables feasible at LHC energies. 
\end{abstract}


\maketitle

\section{Introduction}

A long-standing puzzle in the particle physics is the nature of the Pomeron ($\pom$). This object, with the vacuum quantum numbers, was introduced phenomenologically in the Regge theory as a simple moving pole in the complex angular momentum plane, to describe the high-energy behaviour of the total and elastic cross-sections of the hadronic reactions \cite{collins}. Due to its zero color charge the Pomeron is associated with diffractive events, characterized by the presence of large rapidity gaps in the hadronic final state. The diffractive processes have attracted much attention as a way of amplifying the physics programme at hadronic colliders, including searching for New Physics (For a recent review see, e.g. Ref. \cite{forshaw}). The investigation of these reactions at high energies gives important information about the structure of hadrons and their interaction mechanisms. 

The diffractive processes can be classified as inclusive or exclusive events (See e.g. \cite{forshaw}).
In exclusive events, empty regions in pseudo-rapidity, called rapidity gaps, separate the intact very forward hadron from the central massive object. Exclusivity means that nothing else is produced except the leading hadrons and the central object. 
The inclusive processes also exhibit rapidity gaps. However, they contain soft particles accompanying the production of a hard diffractive object, with the rapidity gaps becoming, in general, smaller than in the exclusive case. Moreover, 
the inclusive  processes can also be classified as single or double diffractive, which is directly associated to the presence of one or two rapidity gaps in the final state, respectively.

The diffractive physics has been tested in hadron-hadron collisions considering  distinct final states like dijets, electroweak vector bosons, dileptons, heavy quarks, quarkonium + photon, and different theoretical approaches (See, e.g., Refs. \cite{Covolan:2002kh,roman,golec2,MMM1,ingelman,golec,quark_photon,schurek,Brodsky:2006wb,Kopeliovich:2006tk,Kopeliovich:2007vs,Pasechnik:2011nw, Kepka:2010hu,Marquet:2012ra}). One of these approaches is the Resolved Pomeron Model,  proposed by Ingelman and Schlein in Ref. \cite{IS}, which assumes the validity of the diffractive factorization formalism and that the Pomeron has a partonic structure. The basic idea is that the hard scattering resolves the quark and gluon content in the Pomeron \cite{IS}, which can be obtained by analysing the experimental data from diffractive deep inelastic scattering (DDIS) at HERA, providing us with the diffractive distributions of singlet quarks and gluons in the Pomeron \cite{H1diff}. However, other approaches based on very distinct assumptions,  for example the BFKL Pomeron, are also able to describe the current scarce experimental data. Consequently, the present scenario for diffractive processes is unclear, motivating the study of alternative processes which could allow to constrain the correct description of the Pomeron.

In Ref. \cite{crisvic_dis} we have proposed, for the first time, the study of the photon production as a complementary test of diffractive processes and the pomeron structure. In this paper we present a detailed analysis of the rapidity and transverse momentum dependence of the cross section for the diffractive double photon and photon+jet production. Moreover, a comparison with the inclusive production is presented. Our motivation is associated to the fact that the production of prompt photons is dominated at leading order by the  QCD Compton subprocess $q g \rightarrow \gamma q$ (See, e.g., \cite{Owens:1986mp}), which makes this process a probe of the diffractive gluon {\it and} quark distributions, in contrast to diffractive heavy quark production which probes mainly the diffractive gluon distribution. 

The content of this paper is as follows. In the next we present a brief review of the formalism for the prompt photon and double photon production in inclusive and diffractive processes. In Section \ref{results} we present our predictions for the rapidity and transverse momentum distributions as well as for the total cross sections and event rates. In Section \ref{conc} we summarize our main conclusions.

\section{Single and Double Diffractive photon production}
\label{sec:single}

In the following we apply the Resolved Pomeron Model \cite{IS} for the diffractive photon production. This model assumes that the Pomeron has a well defined  partonic structure and that the hard process takes place in a Pomeron - proton or proton - Pomeron (single diffractive), or Pomeron - Pomeron (double diffractive) processes. 
At leading order the prompt photon production the hard process is determined by 
 the Compton process $qg \rightarrow q \gamma$,  the annihilation process  $q \bar q  \rightarrow g \gamma$ and the processes  $q \bar q  \rightarrow \gamma \gamma$ (pure EM), $g g  \rightarrow \gamma \gamma$ and $g g  \rightarrow g \gamma$ \cite{Owens:1986mp}. Higher order contributions are not considered here and can be taken into account effectively with a $K$ factor.
In order to estimate the hadronic cross sections we have to convolute the cross sections for these partonic subprocesses with the inclusive and/or diffractive parton distribution functions, depending on the process under analysis.

\begin{figure}[t]
\begin{center}
\scalebox{0.30}{\includegraphics{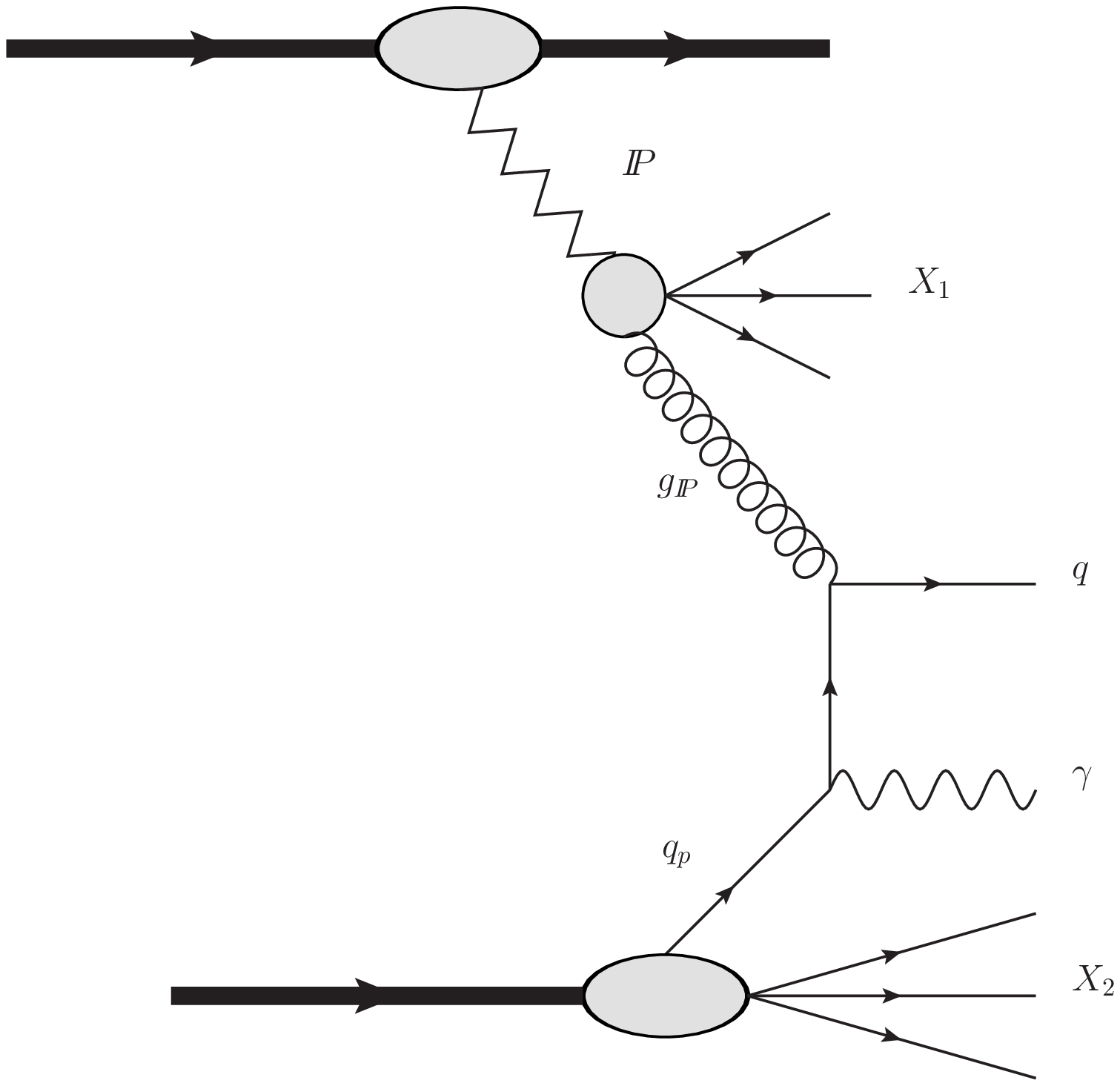}}
\scalebox{0.25}{\includegraphics{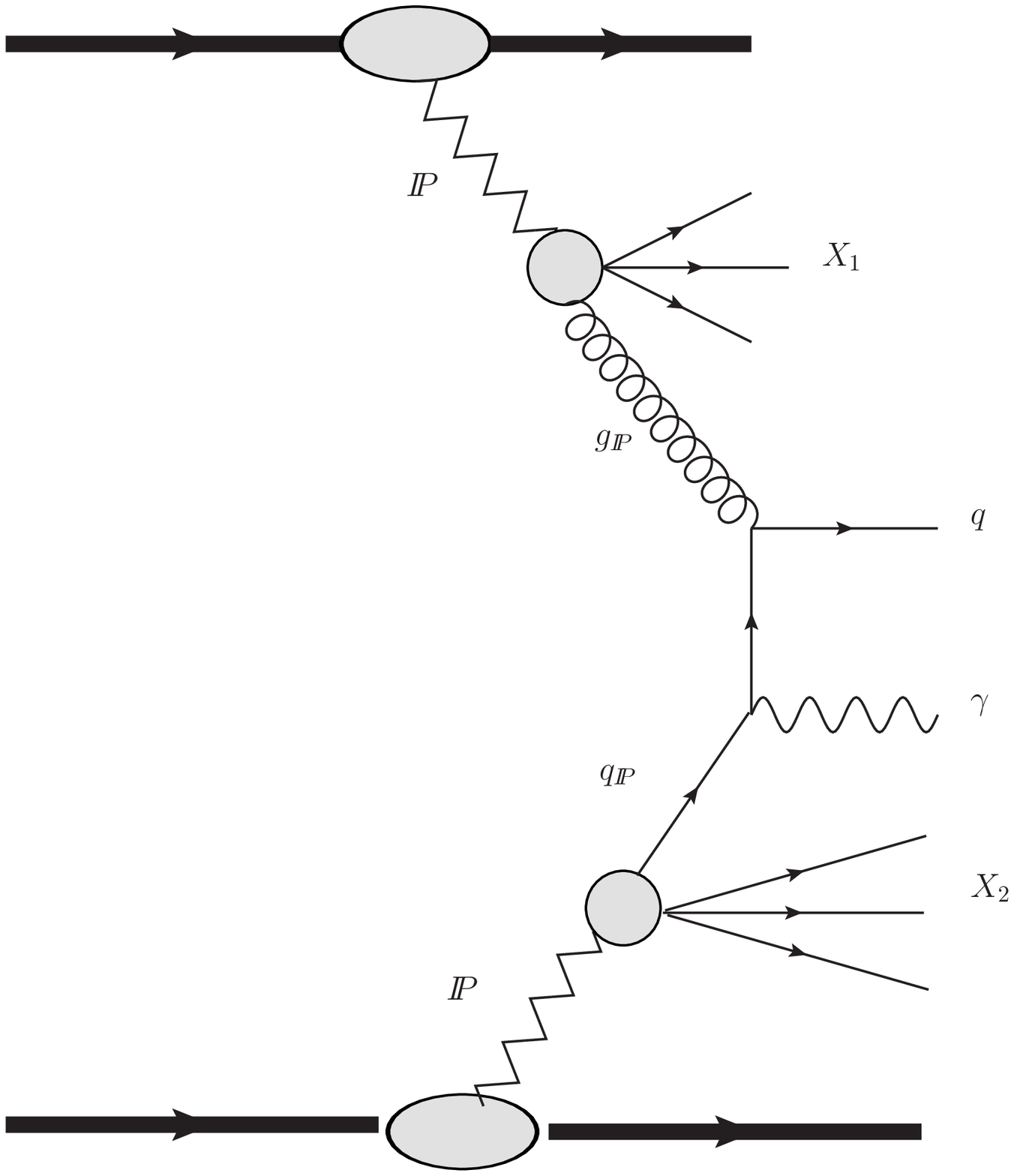}}
\caption{Photon + jet production in single (left panel) and double (right panel) diffractive processes.}
\label{single}
\end{center}
\end{figure}

The photon + jet production in single and double diffractive processes is described by the diagrams like those presented in Fig. 
\ref{single}. Moreover, we include the contributions associated to the production of a photon and an unobserved photon ($\gamma X$). In single prompt photon diffractive production the Pomeron might be emitted from one of the two protons and one should include both $pI\!\!P$ and $I\!\!Pp$ interactions. The final state will be characterized by the presence of one rapidity gap. The corresponding cross section may be written as (See e.g. Ref. \cite{crisvic_photon})
 \begin{eqnarray}
\frac{d\sigma }{dydp_T^2}= \sum_{abcd} \int_{x_{a\, min}}^1 dx_a
\left[f^D_a(x_a,Q^2) f_b(x_b,Q^2) + f_a(x_a,Q^2)f^D_b(x_b,Q^2)\right]
\frac{x_ax_b}{2x_a-x_Te^y}\frac{d\hat{\sigma}}{d\hat{t}}(ab\rightarrow
cd)
\end{eqnarray}
where $f_i(x_i,Q^2)$ and $f_i^D(x_i,Q^2)$ are the inclusive  and  diffractive parton distribution functions, respectively. 
The momentum fraction variables are $x_{a}$ and
$x_b=\frac{x_ax_Te^{-y}}{2x_a-x_Te^y}$, with $x_{a\, min}=\frac{x_Te^{y}}{2-x_Te^{-y} }$ and $x_T=2p_T/\sqrt{s}$. $\frac{d\hat{\sigma}}{d\hat{t}}$ are the LO partonic cross sections for all considered subprocesses.
In the case of double diffractive prompt photon production (also called central diffractive)  one has  ${I\!\!P} {I\!\!P} $ interactions and one looks for events with two rapidity gaps and a $\gamma +$jet at the central region. The corresponding cross section is given by
 \begin{eqnarray}
\frac{d\sigma }{dydp_T^2}= \sum_{abcd} \int_{x_{a\, min}}^1 dx_a
f^D_a(x_a,Q^2) f^D_b(x_b,Q^2)
\frac{x_ax_b}{2x_a-x_Te^y}\frac{d\hat{\sigma}}{d\hat{t}}(ab\rightarrow
cd)\,\,,
\end{eqnarray}
 which is  strongly sensitive to the Pomeron structure due to the quadratic dependence on the diffractive parton distributions. 

\begin{figure}[t]
\begin{center}
\scalebox{0.30}{\includegraphics{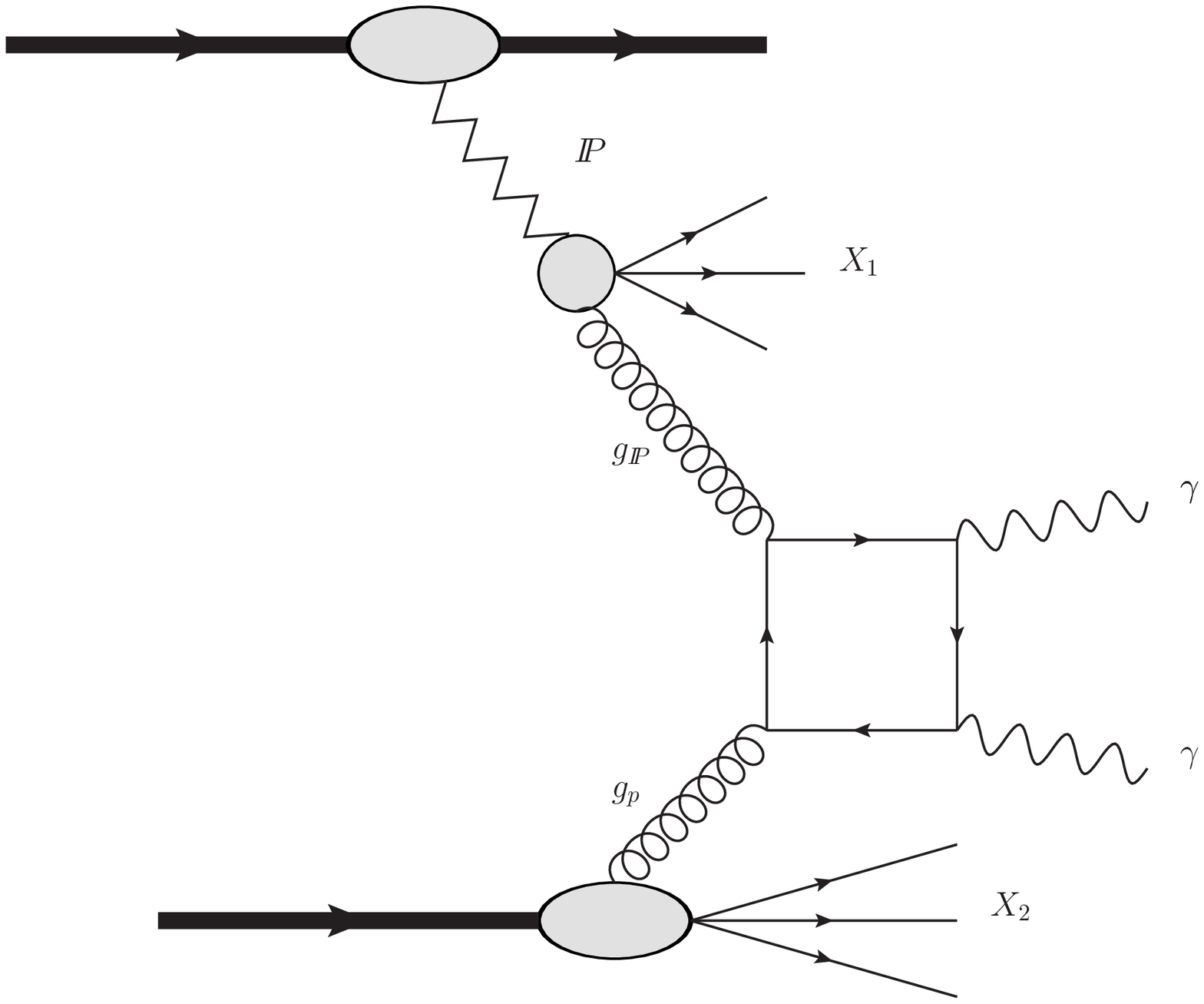}}
\scalebox{0.25}{\includegraphics{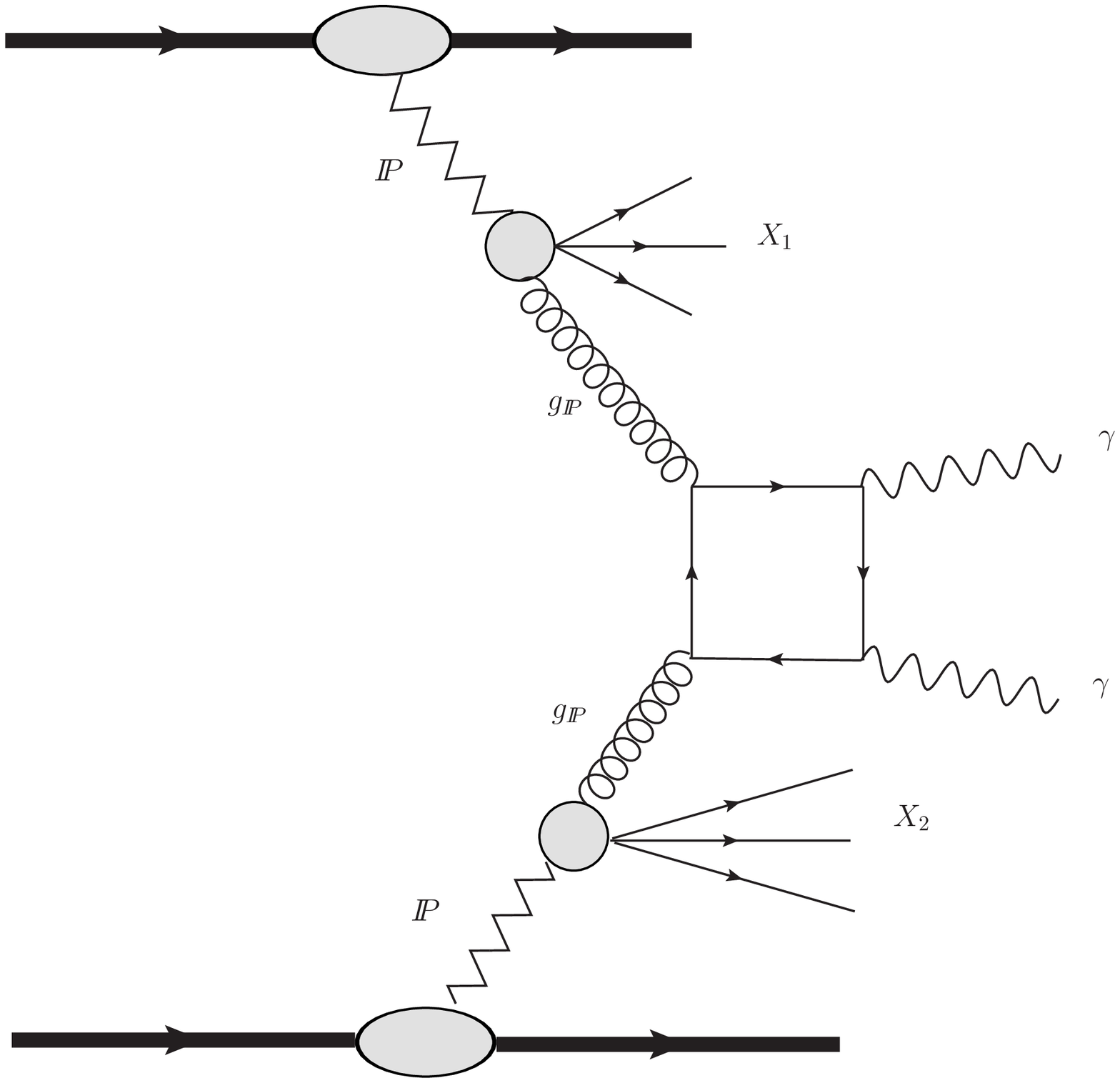}}
\caption{Double photon production in single (left panel) and double (right panel) diffractive processes.}
\label{figsdddouble}
\end{center}
\end{figure}

The double photon production in single and double diffractive processes are described by the diagrams like those shown in Fig. \ref{figsdddouble}, taking into account the $q \bar q  \rightarrow \gamma \gamma$ and $g g  \rightarrow \gamma \gamma$ subprocesses. Besides the rapidity and transverse momentum distributions given above it is useful in the double photon production to analyze the distribution on the invariant mass of the $\gamma \gamma$ system, for which the single diffractive processes can be written as
 \begin{eqnarray}
\frac{d\sigma }{dy_1dy_2dM_{\gamma \gamma}}= \sum_{ab}
\left[ f^D_a(x_a,Q^2)f_b(x_b,Q^2) + f_a(x_a,Q^2) f^D_b(x_b,Q^2)\right]
\frac{x_ax_b M_{\gamma \gamma}}{1+\cosh(y_1-y_2)}\frac{d\hat{\sigma}}{d\hat{t}}(ab\rightarrow
\gamma \gamma)\,,
\label{mass}
\end{eqnarray}
where $x_{a}=\frac{p_T}{\sqrt{s}}[e^{y_1}+e^{y_2}]$, $x_{b}=\frac{p_T}{\sqrt{s}}[e^{-y_1}+e^{-y_2}]$, and $y_1$, $y_2$ are the rapidities of the outgoing photons. For double diffractive processes, the factor in the brackets in Eq. (\ref{mass}) is replaced by $f^D_a(x_a,Q^2) f^D_b(x_b,Q^2)$.

In the present work, the diffractive parton distributions in the proton are taken from the Resolved Pomeron Model \cite{IS}, where they are defined as a convolution of the Pomeron flux emitted by the proton, $f_{I\!\!P}(x_{I\!\!P})$, and the parton distributions in the Pomeron, $g_{I\!\!P}(\beta, \mu^2)$, $q_{I\!\!P}(\beta, \mu^2)$, where $\beta$ is the momentum fraction carried by the partons inside the Pomeron. 
The Pomeron flux is given by $f_{I\!\!P}(x_{I\!\!P})= \int_{t_{min}}^{t_{max}} dt f_{\pom/p}(x_{{I\!\!P}}, t)$, where $f_{\pom/p}(x_{\pom}, t) = A_{\pom} \cdot \frac{e^{B_{\pom} t}}{x_{\pom}^{2\alpha_{\pom} (t)-1}}$ and $t_{min}$, $t_{max}$ are kinematic boundaries. The Pomeron flux factor is motivated by Regge theory, where the Pomeron trajectory assumed to be linear, $\alpha_{\pom} (t)= \alpha_{\pom} (0) + \alpha_{\pom}^\prime t$, and the parameters $B_{\pom}$, $\alpha_{\pom}^\prime$ and their uncertainties are obtained from fits to H1 data  \cite{H1diff}. 
The diffractive quark and gluon distributions are then given by
\begin{eqnarray}
{ q^D(x,\mu^2)}=\int dx_{I\!\!P}d\beta \delta (x-x_{I\!\!P}\beta)f_{I\!\!P}(x_{I\!\!P})q_{I\!\!P}(\beta, \mu^2)={ \int_x^1 \frac{dx_{I\!\!P}}{x_{I\!\!P}} f_{I\!\!P}(x_{I\!\!P}) q_{I\!\!P}\left(\frac{x}{x_{I\!\!P}}, \mu^2\right)} \\
{ g^D(x,\mu^2)}=\int dx_{I\!\!P}d\beta \delta (x-x_{I\!\!P}\beta)f_{I\!\!P}(x_{I\!\!P})g_{I\!\!P}(\beta, \mu^2)={ \int_x^1 \frac{dx_{I\!\!P}}{x_{I\!\!P}} f_{I\!\!P}(x_{I\!\!P}) g_{I\!\!P}\left(\frac{x}{x_{I\!\!P}}, \mu^2\right)}
\end{eqnarray}
In our analysis we use the diffractive parton distributions obtained by the H1 Collaboration at DESY-HERA \cite{H1diff}. Moreover, we use the inclusive parton distributions as given by the CTEQ6L parametrization \cite{Pumplin:2002vw}.

In order to obtain reliable predictions for the diffractive cross sections one should take into account that the QCD hard scattering factorization theorem for diffraction is violated in $pp$ collisions by soft interactions which lead to an extra production of particles that destroy the rapidity gaps related to pomeron exchange. The inclusion of these additional absorption effects can be parametrized in terms of a rapidity gap survival probability, $S^2$, which corresponds to the probability of the scattered proton not to dissociate due to the secondary interactions. These effects have been calculated considering different approaches giving distinct predictions (See, e.g. Ref. \cite{review_martin}). An usual approach in the literature is the calculation of an average probability  
$\langle |S|^2\rangle$ and after to multiply  the cross section by this value. As previous studies for single and double diffractive production \cite{MMM1,schurek} we also follow this simplified approach assuming $\langle |S|^2\rangle = 0.05$ for single diffractive processes and  $\langle |S|^2\rangle = 0.02$ for double diffractive processes. These values are taken from Ref. \cite{KMR}.

\section{Results}
\label{results}

\begin{figure}[t]
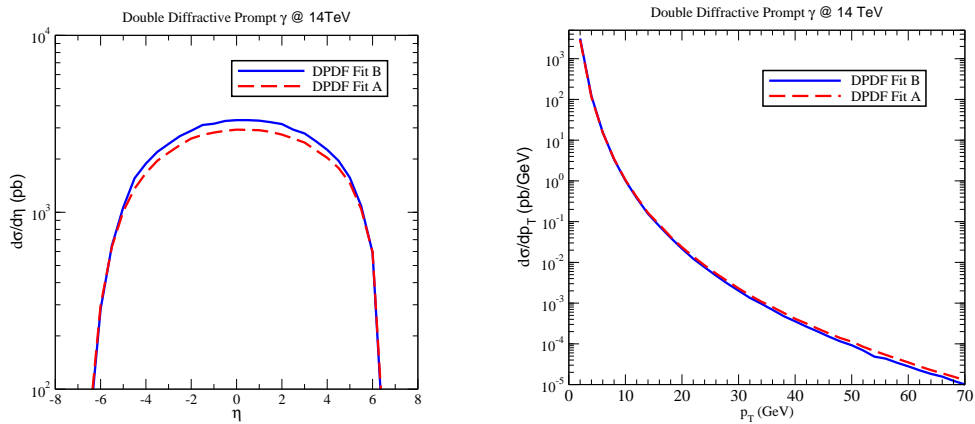

\begin{center}
\scalebox{0.42}{\includegraphics{ddpromptgamma_eta_h1fits.eps}}
\hspace{1cm}
\scalebox{0.42}{\includegraphics{ddpromptgamma_pt_h1fits.eps}}
\caption{(Color online). Dependence on the diffractive parton distributions of the  pseudorapidity (left panel) and transverse momentum (right panel) distributions  for the double diffractive prompt photon production at $\sqrt{s}= 14$ TeV.}
\label{h1fits}
\end{center}
\end{figure}

In what follows we present our results for the diffractive photon production at LHC energies. Initially we analyse the dependence of our predictions on the diffractive parton distribution used in the calculations. The H1 Collaboration  proposed two different parametrizations for the diffractive parton distributions (denoted A and B in Ref. \cite{H1diff}), which differ in the form of the parton densities at the starting scale for the QCD evolution and in the value of the effective pomeron intercept, being   $\alpha_{\pom} (0)=1.118\pm 0.008$ in fit A and $\alpha_{\pom} (0)=1.111\pm 0.007$ in fit B.
In Fig. \ref{h1fits} (left panel) we present our results for the rapidity distribution for the photon+jet production in double diffractive processes (denoted Double Diffractive Prompt $\gamma$ hereafter). We consider this process  for this analysis due to the quadratic  dependence on $f^D$.  The difference between the predictions is  small ($\lesssim 10$\%), mainly in $|\eta|<4$ region. Similar results are obtained for the transverse momentum dependence, with the difference being at large-$p_T$ [See Fig. \ref{h1fits} (right panel)]. In the following calculations we will use the diffractive parton distributions as given by the fit B in Ref. \cite{H1diff}.

\begin{figure}[t]
\begin{center}
\scalebox{0.4}{\includegraphics{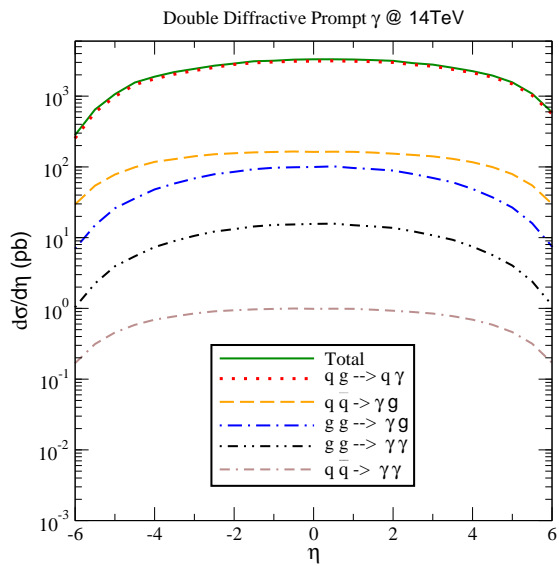}}
\caption{(Color online). Contribution of the different partonic subprocesses for the double diffractive prompt photon production  at $\sqrt{s}= 14$ TeV.}
\label{canais}
\end{center}
\end{figure}

Lets analyse now the contribution of the different partonic subprocesses for the double diffractive prompt $\gamma$ production. In Fig. \ref{canais} the distinct contributions are explicitly presented. As expected from the inclusive case, the cross section for this process is dominated by the QCD Compton subprocess $qg \rightarrow q \gamma$, with the annihilation channel $q \bar q  \rightarrow g \gamma$ being smaller by one order of magnitude. Moreover,  the double photon production is dominated by the $g g  \rightarrow \gamma \gamma$ subprocess.

\begin{figure}[t]
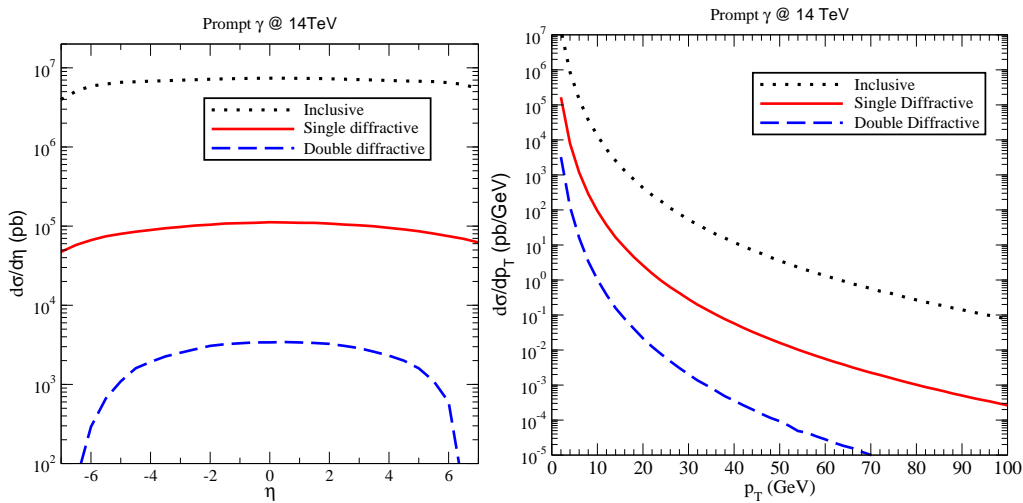

\begin{center}
\scalebox{0.35}{\includegraphics{promptgamma_eta.eps}}
\scalebox{0.35}{\includegraphics{promptgamma_pt.eps}}
\caption{(Color online). Pseudorapidity (left panel) and transverse momentum (right panel) distributions for the prompt photon production at $\sqrt{s}= 14$ TeV.}
\label{fig:prompt}
\end{center}
\end{figure}

In Fig. \ref{fig:prompt} we present our predictions for the prompt photon production in single and double diffractive processes. 
For the sake of comparison, we also show the inclusive (non-diffractive) LO predictions. One observes a reduction of two orders of magnitude
when going from the inclusive to single diffractive, and from single to double diffractive. However, as verified in Table \ref{tab1}, the diffractive cross sections and event rates/second are still sizeable and could be measured at the LHC. In our calculations of the event rates we have assumed 
the  design luminosity ${\cal L}^{\mathrm{pp}}_{\mathrm{LHC}} = 10^7$ mb$^{-1}$s$^{-1}$.

\begin{figure}[t]
\begin{center}
\scalebox{0.49}{\includegraphics{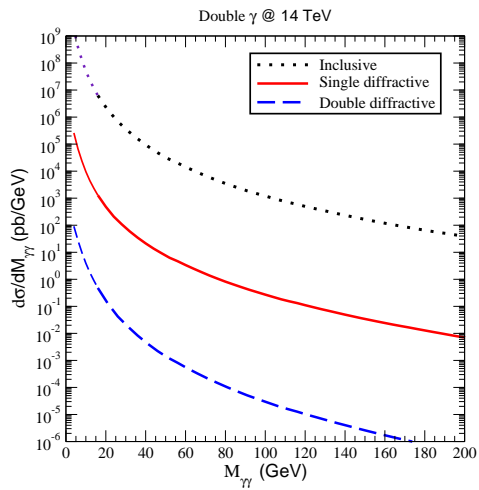}}
\caption{(Color online). Distributions on the invariant mass of the $\gamma \gamma$ system, in double photon production at $\sqrt{s}= 14$ TeV for inclusive, single and central diffractive processes. The rapidity of the photons is integrated in the range $|y_{1,2}|\le 8$.}
\label{fig:doubleM}
\end{center}
\end{figure}

The results for double photon production are shown in Figs. \ref{fig:double} and \ref{fig:doubleM}. Compared with prompt photon production, one has much cleaner processes because of the absence of a produced jet. On the other hand, the cross sections are much smaller (by three orders of magnitude), since only the subdominant subprocesses contribute. On the single diffractive case, one could in principle measure photon pairs with invariant mass up to higher values. For the double diffractive case, the distribution drops fast and then only lower values for the diphoton invariant mass would be accessible. The total cross sections and event rates/second are presented in Table \ref{tab1}.
 
\begin{table}[h]
\begin{center}
 \vspace{0.5cm}
\begin{tabular}{|c|c|c|c|}
\hline
\hline
Final state & Inclusive & Single diffractive & Double diffractive  \\
\hline
\hline
 $\gamma $ jet & $10169 \times 10^4$ pb ( $1016900$) & 137$\times 10^4$ pb (13700)  &  2.95$\times 10^4$ pb (295)  \\
\hline
 $\gamma \gamma$ & $2.98\times 10^5$ pb ($2980$) & 4617.0 pb (46.2)  & 128.0  pb (1.28)  \\
  \hline
\hline
\end{tabular}
\caption{Total cross sections and event rates/second for photon+jet and double photon production.}
\label{tab1}
\end{center}
\end{table}

\begin{figure}[t]
\begin{center}
\scalebox{0.3}{\includegraphics{doublegamma_etav2.eps}}
\hspace{1cm} 
\scalebox{0.3}{\includegraphics{doublegamma_ptv2.eps}}
\caption{(Color online). Pseudorapidity (left panel) and transverse momentum (right panel) distributions for the double photon production at $\sqrt{s}= 14$ TeV.}\label{fig:double}
\end{center}
\end{figure}

Some comments are in order. Firstly, the double photon production in central exclusive diffractive processes, $pp \rightarrow p \gamma \gamma p$, where nothing else is produced except the leading hadrons and the two photons, have been studied in \cite{kmrs}, observed in $p\bar{p}$ collisions at Tevatron \cite{tevatron} and searched for at the LHC \cite{lhc_photons}. At $\sqrt{s} = 14$ TeV, the cross section is predicted to be equal to  20 pb for diphoton masses $M_{\gamma \gamma} > 5$ GeV \cite{kmrs}. Moreover, the exclusive double photon production in $pp$ collisions was recently studied in Ref. \cite{gustavo} considering the incident protons as photon emitters and the elastic light-by-light scattering ($\gamma \gamma \rightarrow \gamma \gamma$), obtaining $\sigma^{excl}_{\gamma \gamma \rightarrow \gamma \gamma} [M_{\gamma \gamma} > 5 \mbox{GeV}] = 12$ fb at $\sqrt{s} = 14$ TeV. In contrast, for the double diffractive $\gamma \gamma$ production considered in this paper, which is characterized by the presence of remnants of the Pomeron beyond the leading hadrons and the two photons, we predict $\sigma^{DD} (M_{\gamma \gamma} > 5 \, \mbox{GeV}) = 120$ pb. If we consider the Tevatron energy and the rapidity range probed in \cite{tevatron}, given by  $|\eta(\gamma)|< 1$, we predict  $\sigma^{DD}$ = 6.6 pb, in contrast with the central exclusive prediction of 0.2 - 2 pb, depending on the unintegrated gluon distribution used in the calculations, which  reasonably describes the experimental data. It is important to emphasize that it is not obvious if the double diffractive and the central exclusive mechanisms could be differentiated experimentally at the LHC. Unfortunately, due to the high luminosity and large pile-up environment the separation of the diffractive processes considering the rapidity gaps and the detection or not of the remnants of the Pomeron will be a hard task. Another comment is that distinctly from the heavy quark and dijet production, which can be produced in exclusive and inclusive double diffractive processes, the production of photon + jet can only occur in the latter case, which makes the experimental analysis of this final state ideal to study the validity of the Resolved Pomeron Model. Moreover, as already emphasized in the Introduction, in the photon + jet production we also are probing the diffractive  quark distributions. Therefore, the combined  analysis of this process and heavy quark or double photon production can be useful to constrain these distributions.

\section{Summary}
\label{conc}

As a summary, in this paper we have presented a detailed analysis  for the diffractive photon production  in $pp$ collisions at the LHC. We have discussed this process in the framework of the Resolved Pomeron model corrected for absorption effects, as used in the estimation of several other diffractive processes.  Our results indicate that the experimental analysis is feasible and it would help in constrain the underlying model for the Pomeron and the diffractive parton distributions.

\begin{acknowledgments}
 This research was supported by CNPq, CAPES and FAPERGS, Brazil. 
\end{acknowledgments}


\begin{thebibliography}{99}

\bibitem{collins} P. D. B. Collins, {\it An Introduction to Regge theory and high energy physics} (Cambridge University Press, Cambridge, England, 1977).


\bibitem{forshaw} 
  M.~G.~Albrow, T.~D.~Coughlin and J.~R.~Forshaw,
  %``Central Exclusive Particle Production at High Energy Hadron Colliders,''
  Prog.\ Part.\ Nucl.\ Phys.\  {\bf 65}, 149 (2010)
  

\bibitem{Covolan:2002kh} 
 R.~J.~M.~Covolan and M.~S.~Soares,
  %``Diffractive hadroproduction of dijets and $W$ ' $s$ at the Tevatron collider and the pomeron structure function,''
  Phys.\ Rev.\ D {\bf 67}, 017503 (2003);  M.~B.~Gay Ducati, M.~M.~Machado and M.~V.~T.~Machado,
  %``Diffractive hadroproduction of $W^\pm$ and $Z^0$ bosons at high energies,''
  Phys.\ Rev.\ D {\bf 75}, 114013 (2007)



\bibitem{roman} 
  R.~Pasechnik, B.~Kopeliovich and I.~Potashnikova,
  %``Diffractive Gauge Bosons Production beyond QCD Factorisation,''
  Phys.\ Rev.\ D {\bf 86}, 114039 (2012)

\bibitem{golec2} 
  K.~Golec-Biernat, C.~Royon, L.~Schoeffel and R.~Staszewski,
  %``Electroweak vector boson production at the LHC as a probe of mechanisms of diffraction,''
  Phys.\ Rev.\ D {\bf 84}, 114006 (2011)

\bibitem{MMM1} M.V.T. Machado, Phys.\ Rev.\  D {\bf 76}, 054006  (2007); M. B. Gay Ducati, M. M. Machado, M. V. T. Machado, Phys.\ Rev.\ {\bf D81}, 054034 (2010); 
  M.~B.~Gay Ducati, M.~M.~Machado and M.~V.~T.~Machado,
  %``Charm and bottom production in inclusive double Pomeron exchange in heavy ion collisions at the LHC,''
  Phys.\ Rev.\ C {\bf 83}, 014903 (2011)   

\bibitem{ingelman} 
  G.~Ingelman, R.~Pasechnik, J.~Rathsman and D.~Werder,
  %``Diffractive W production at hadron colliders as a test of colour singlet exchange mechanisms,''
  Phys.\ Rev.\ D {\bf 87}, 094017 (2013)

\bibitem{golec} 
  K.~Golec-Biernat and A.~Luszczak,
  %``Diffractive production of electroweak vector bosons at the LHC,''
  Phys.\ Rev.\ D {\bf 81}, 014009 (2010)

  
\bibitem{quark_photon} 
  M.~B.~Gay Ducati, M.~M.~Machado and M.~V.~T.~Machado,
  %``Diffractive quarkonium production in association with a photon at the LHC,''
  Phys.\ Lett.\ B {\bf 683}, 150 (2010)  

\bibitem{schurek} 
G. Kubasiak and A.~Szczurek,
  Phys.\ Rev.\ D {\bf 84}, 014005 (2011)

\bibitem{Brodsky:2006wb} 
  S.~J.~Brodsky, B.~Kopeliovich, I.~Schmidt and J.~Soffer,
  %``Diffractive Higgs production from intrinsic heavy flavors in the proton,''
  Phys.\ Rev.\ D {\bf 73}, 113005 (2006)

\bibitem{Kopeliovich:2006tk} 
  B.~Z.~Kopeliovich, I.~K.~Potashnikova, I.~Schmidt and A.~V.~Tarasov,
  %``Unusual features of Drell-Yan diffraction,''
  Phys.\ Rev.\ D {\bf 74}, 114024 (2006)
  
\bibitem{Kopeliovich:2007vs} 
  B.~Z.~Kopeliovich, I.~K.~Potashnikova, I.~Schmidt and A.~V.~Tarasov,
  %``Diffractive Excitation of Heavy Flavors: Leading Twist Mechanisms,''
  Phys.\ Rev.\ D {\bf 76}, 034019 (2007)

\bibitem{Pasechnik:2011nw} 
  R.~S.~Pasechnik and B.~Z.~Kopeliovich,
  %``Drell-Yan diffraction: Breakdown of QCD factorisation,''
  Eur.\ Phys.\ J.\ C {\bf 71}, 1827 (2011)

\bibitem{Kepka:2010hu} 
  O.~Kepka, C.~Marquet and C.~Royon,
  %``Gaps between jets in hadronic collisions,''
  Phys.\ Rev.\ D {\bf 83}, 034036 (2011)


\bibitem{Marquet:2012ra} 
  C.~Marquet, C.~Royon, M.~Trzebiński and R.~Žlebčík,
  %``Gaps between jets in double-Pomeron-exchange processes at the LHC,''
  Phys.\ Rev.\ D {\bf 87},  034010 (2013)  

\bibitem{IS} G. Ingelman and P.E. Schlein, Phys. Lett. {\bf B152}, 256 (1985).

\bibitem{H1diff} H1 Collab.,  A. Aktas {\it et al.}, Eur. Phys. J. {\bf C48}, 715 (2006).


\bibitem{crisvic_dis}
  C.~Brenner Mariotto and V.~P.~Goncalves, {\it Single and Double diffractive prompt photon production at the LHC}. Talk presented to the  XXI International Workshop on Deep-Inelastic Scattering and Related Subjects, 22 - 26 April 2013, Marseille, France. To be published in the Proceedings of Science (PoS). 

\bibitem{Owens:1986mp} 
  J.~F.~Owens,
  %``Large Momentum Transfer Production of Direct Photons, Jets, and Particles,''
  Rev.\ Mod.\ Phys.\  {\bf 59}, 465 (1987).

\bibitem{crisvic_photon}
  C.~Brenner Mariotto and V.~P.~Goncalves,
  %``Enhancement of prompt photons in ultrarelativistic proton-proton collisions from nonlinear gluon evolution at small-x,''
  Phys.\ Rev.\ C {\bf 75}, 068202 (2007) 


\bibitem{Pumplin:2002vw}
  J.~Pumplin, D.~R.~Stump, J.~Huston, H.~L.~Lai, P.~Nadolsky and W.~K.~Tung,
  JHEP {\bf 0207}, 012 (2002).



\bibitem{review_martin} 
  M.~G.~Ryskin, A.~D.~Martin, V.~A.~Khoze and A.~G.~Shuvaev,
  %``Soft physics at the LHC,''
  J.\ Phys.\ G {\bf 36}, 093001 (2009)


\bibitem{KMR} V.A. Khoze,  A.D. Martin, M.G. Ryskin, Eur. Phys. J. {\bf C18}, 167 (2000).

  

\bibitem{kmrs} 
  V.~A.~Khoze, A.~D.~Martin, M.~G.~Ryskin and W.~J.~Stirling,
  %``Diffractive gamma-gamma production at hadron colliders,''
  Eur.\ Phys.\ J.\ C {\bf 38}, 475 (2005)
  
\bibitem{tevatron} 
  T.~Aaltonen {\it et al.}  [CDF Collaboration],
  %``Observation of Exclusive Gamma Gamma Production in $p \bar{p}$ Collisions at $\sqrt{s}=1.96$ TeV,''
  Phys.\ Rev.\ Lett.\  {\bf 108}, 081801 (2012)

\bibitem{lhc_photons} 
  S.~Chatrchyan {\it et al.}  [CMS Collaboration],
  %``Search for exclusive or semi-exclusive photon pair production and observation of exclusive and semi-exclusive electron pair production in $pp$ collisions at $\sqrt{s}=7$ TeV,''
  JHEP {\bf 1211}, 080 (2012)
  
\bibitem{gustavo}
D. d'Enterria and G. G. da Silveira, Phys.\ Rev.\ Lett. {\bf 111}, 080405 (2013).   



    
\end{thebibliography}
\end{document}